\documentclass{ws-spin}
\usepackage{multicol}
\usepackage{bm,color}
\usepackage[mathscr]{eucal}
\usepackage[breaklinks=true,unicode=true,urlcolor = blue,colorlinks = true,citecolor = blue,linkcolor = blue]{hyperref}
\usepackage[numbers,sort&compress]{natbib}
\usepackage{graphicx}
\usepackage[T1]{fontenc}
\usepackage{pgf}
\usepackage{tikz}
\usetikzlibrary{calc,intersections}
\renewcommand{\vec}[1]{\bm{#1}}
\begin{document}

\markboth{D.~D.~Sheka, V.~P.~Kravchuk, M.~I.~Sloika, Yu.~Gaididei}{Equilibrium states of soft magnetic hemispherical shell}

\title{EQUILIBRIUM STATES OF SOFT MAGNETIC HEMISPHERICAL SHELL}

\author{DENIS D. SHEKA$^{1,*}$, VOLODYMYR P. KRAVCHUK$^{2,\dag}$, MYKOLA I. SLOIKA$^{1,\ddag}$, YURI GAIDIDEI$^{2,\S}$}

\address{$^1$ Taras Shevchenko National University of Kiev, 01601 Kiev, Ukraine\\
$^2$ Bogolyubov Institute for Theoretical Physics, 03143 Kiev, Ukraine\\
$^*$\email{sheka@univ.net.ua}
$^\dag$\email{vkravchuk@bitp.kiev.ua}
$^\ddag$\email{sloika.m@gmail.com}
$^\S$\email{ybg@bitp.kiev.ua}
}

\maketitle

\begin{history}
\received{June 5 2013}
\revised{August 5 2013}
\end{history}

\begin{abstract}
The ground state of hemispherical permalloy magnetic shell is studied. There exist two magnetic phases: the onion state and the vortex one. The phase diagram is systematically analyzed in a wide range of geometrical parameters.  Possible transitions between different phases are analyzed using the combination of analytical calculations and micromagnetic simulations.
\end{abstract}

\keywords{micromagnetism; magnetic nanoparticle; vortex state; onion state}

\begin{multicols}{2}

\section{Introduction}

During last years there appears a growing interest in studying of topological structures on curved surfaces. This interest becomes stronger on the one hand due to the interplay between geometry, curvature and physical properties of topological structures. On the other hand, recent developments in nanotechnology makes it possible to produce nanoparticles of various shapes. For instance, in nanomagnetism, topologically nontrivial structures, magnetic vortices, can form ground states of nanosamples \cite{Guimaraes09}. The control vortex statics and dynamics in scales of nanometers and picoseconds is of crucial importance for applications such as magnetic logic and memory concepts \cite{Bohlens08,Pigeau10,Yu11a,Hertel13,Uhlir13}.

During last decade different aspects of magnetic vortex statics and dynamics were studied; however, investigations were mainly restricted to flat geometry. It is well known that the vortex state appears as a ground state in sub--micrometer--sized magnets due to competition between short--range exchange interactions and the long--range dipole interactions \cite{Hubert98}. For the smaller samples the ground state is the monodomain one, which is characterized by the quasiuniform magnetization structure. The curvature can crucially change the physical picture. For example, in spherical shell the quasiuniform in--surface state is forbidden for topological reasons; instead vortex singularities appear. Recently we found that the curvature results in coupling between the localized out--of--surface component of the vortex with its delocalized in--surface structure \cite{Kravchuk12a}. Recent studies of the equilibrium state in soft magnetic permalloy caps on self--assembled spherical particles revealed  the vortex
ground state for individual caps \cite{Streubel12} and  closely packed cap arrays \cite{Streubel12a}. Depending on the geometrical and magnetic parameters of the caps, there exist different magnetization configurations (magnetic phases) in extruded hemispheres: two monodomain phases (the uniform easy--axis phase and the onion one), and the vortex phase \cite{Streubel12}.

In this paper we consider another geometry of the cap, namely a hemispherical shell.
By combining analytical methods and micromagnetic simulations we describe the equilibrium magnetic phases: the monodomain state and the vortex one. The key moment is to propose a simple analytical approach, which allows us to describe different states and transitions between them. We will see below that the geometry of hemispherical shell results in a significantly modified phase diagram without a uniform easy--axis state in comparison with extruded hemispheres.

\section{The model. Uniform state}
\label{sec:model}

We consider a classical magnetically soft particle, using the continuous description for the magnetization. In term of angular variables, the magnetization, normalized by its saturation value $M_s$ has the following form
\begin{equation} \label{eq:m}
\vec{m} = \frac{\vec{M}}{M_s} = \left(\sin\Theta\cos\Phi, \sin\Theta\sin\Phi, \cos\Theta \right).
\end{equation}
Our consideration is limited by two main contributions to the energy functional, the exchange energy $E^{\text{ex}}$ and the magnetostatic energy $E^{\text{ms}}$. We neglect the anisotropy energy contributions, what is a reasonable for the soft materials like permalloy
($\text{Ni}_{80}\text{Fe}_{20}$, Py).

\begin{figurehere}
\begin{center}
\begin{tikzpicture}[scale=1.37]

\coordinate (P1) at (17,14);  
\coordinate (O1) at (0,0);    
\coordinate (N1) at (0,2.05); 
\coordinate (F1) at (-2.3,0);
\coordinate (F2) at  (2.2,0);
\coordinate (A1) at (intersection cs: first line={(F1) -- (P1)},
second line={(-2,-0.5)--(2,-0.5)});
\coordinate (A2) at (intersection cs: first line={(F2) -- (P1)},
second line={(-2,-0.5)--(2,-0.5)});
\coordinate (A3) at (intersection cs: first line={(F1) -- (P1)},
second line={(-2,0.5)--(2,0.5)});
\coordinate (A4) at (intersection cs: first line={(F2) -- (P1)},
second line={(-2,0.5)--(2,0.5)});
\coordinate (A5) at (intersection cs: first line={(O1) -- (P1)},
second line={(A1)--(A2)});
\coordinate (A6) at (intersection cs: first line={(O1) -- (P1)},
second line={(A3)--(A4)});
\coordinate (A7) at (intersection cs: first line={(N1) -- (P1)},
second line={(A5)--($(A5)+(0,3)$)});
\coordinate (A8) at (intersection cs: first line={(N1) -- (P1)},
second line={(A6)--($(A6)+(0,3)$)});
\shadedraw[ball color=cyan!50!magenta]   (2,0) arc (0:180:2);
\filldraw[color=gray!50] (O1) circle (2 and 0.5);
\shadedraw[very thin,ball color=cyan!50!magenta,opacity=0.25]   (1,0) arc (0:180:1);
\shadedraw[left color=white,right color=cyan!50!magenta,draw=gray,name path=ellipse] (O1) circle (1 and 0.25);

\draw[dashed,color=black] (-2,0) -- (-2,-1.2);
\draw[dashed,color=black] (-1,0) -- (-1,-1.2);
\draw[dashed,color=black] (1,0) -- (1,-1.2);
\draw[color=black,<->,thick] (-1.97,-0.95) -- node[above] {$h$} (-1.03,-0.95);
\draw[color=black,<->,thick] (-0.97,-0.95) -- node[above] {$2R$} (0.97,-0.95);
\draw[very thin, fill=green,opacity=0.137] (A1) to (A2) to (A4) to (A3);
\draw[very thin, fill=yellow,opacity=0.2, name path = yellowrect] (A5) to (A6) to (A8) to (A7);
\path [name intersections={of=ellipse and yellowrect}];
\coordinate (A9) at (intersection-1);
\coordinate (A10) at (intersection-2);
%
%
\draw[very thin,color=black, fill=yellow!50!white,opacity=0.25] (A6) arc (10:180:0.61 and 1.82) --  (A5) -- (A10) arc (180:23:0.3 and 1.23) -- (A9) -- cycle;

\coordinate (C0) at (-2.5,1.25);
\coordinate (Cx) at ($(C0)+(0.75,0)$);
\coordinate (Cz) at ($(C0)+(0,0.75)$);
\coordinate (Cy) at (intersection cs: first line={(Cx) -- (Cz)},second line={(C0)--(P1)});
\draw[color=red,->] (C0) --  (Cx) node[above] {\small $x$};
\draw[color=red,->] (C0) --  (Cy) node[above] {\small $y$};
\draw[color=red,->] (C0) --  (Cz) node[above] {\small $z$};
%


\end{tikzpicture}

\end{center}
\caption{Schematic of the hemispherical shell.}
\label{fig:mag-full}
\end{figurehere}

We consider a hemispherical shell of the inner radius $R$ and the shell thickness $h$, see Fig.~\ref{fig:mag-full}. The bottom cross-section coincides with the x--y plane. For the homogeneous magnetization distribution the total energy contains only the contribution of surface magnetostatic charges $\sigma(\vec{r}) = \vec{m}\cdot\vec{n}$ with $\vec{n}$ being external normal to the surface, which consists of two hemispherical surfaces and a bottom ring, see Fig.~\ref{fig:mag-full}.  The demagnetization factor of the homogeneously magnetized sample reads:
\begin{equation} \label{eq:N}
N = \frac{1}{4\pi V} \int_S \int_{S'}
\frac{\sigma(\vec{r}) \sigma(\vec{r'}) \mathrm{d}S \mathrm{d}S'}{|\vec{r} - \vec{r'}|}.
\end{equation}
Let us consider the shell, which is homogeneously magnetized along $x$--axis. In this case $\sigma(\vec r) = \sin\vartheta \cos\varphi$, where we use the spherical reference frame for $\vec{r} = \left(r,\vartheta,\varphi\right)$. The magnetic charges appear only on spherical surfaces, hence one can calculate the demagnetization factor $N_x$ in the following form:
\begin{equation*}
N_x = \mathscr{N}_x(S_{\text{in}};S_{\text{in}}) + \mathscr{N}_x(S_{\text{out}};S_{\text{out}}) - 2 \mathscr{N}_x(S_{\text{in}};S_{\text{out}}),
\end{equation*}
\begin{equation} \label{eq:Nx-full-gen} %
\begin{split}
\mathscr{N}_x(S;S') &= \frac{r^2 {r'}^2}{4\pi V} \int_0^{\pi/2}\!\!\! \sin^2\vartheta \mathrm{d}\vartheta \int_0^{\pi/2}\!\!\! \sin^2\vartheta' \mathrm{d}\vartheta'\\
&\times \int_0^{2\pi} \!\!\! \mathrm{d}\varphi \int_0^{2\pi}\!\!\! \mathrm{d}\varphi'\, \frac{\cos\varphi \cos\varphi'}{\left| \vec{r} - \vec{r}'\right|}.
\end{split}
\end{equation}
One can calculate demagnetization factors using an expansion of $1/|\vec{r}-\vec{r}'|$ over associated Legendre functions \cite{Abramowitz72}
\begin{equation*} \label{eq:expansion-on-Plm}
\begin{split}
\frac{1}{|\vec{r} - \vec{r}'|}  &= \frac{1}{r_>}\sum_{l=0}^\infty\! \sum_{m=-l}^l \left(\frac{r_<}{r_>}\right)^{l} \frac{(l-m)!}{(l+m)!}\\
&\times P_l^m(\cos\vartheta) P_l^m(\cos\vartheta') e^{i m(\varphi-\varphi')}.
\end{split}
\end{equation*}
Here $r_<=\min(r,r')$, $r_>=\max(r,r')$, and $P_l^m(z)$ is the associated Legendre polynomial. Then one can rewrite \eqref{eq:Nx-full-gen} as follows:
\begin{align} \label{eq:Nx-full-2}
&N_x = \frac{3}{4(\varkappa^3-1)}\left[(1+\varkappa^3)\mathcal{S}(1) - 2 \mathcal{S}(\varkappa) \right],\nonumber \\
&\mathcal{S}(x) = \sum_{l=0}^\infty \frac{s_l^2}{l(l+1) x^{l-1}}, \qquad \varkappa = 1+\varepsilon,\\
&s_l = \int_0^{\pi/2} \!\!\!\sin^2\vartheta P_l^1(\cos\vartheta) \mathrm{d}\vartheta = \frac{\sqrt{\pi} l(l+1)}{4 \Gamma\left( \frac32-\frac{l}2\right) \Gamma\left( 2+ \frac{l}2\right) },\nonumber
\end{align}
where $\varepsilon=h/R$ is an aspect ratio and $\Gamma(x)$ is Euler's gamma--function. Using the explicit form of $s_l$, the series can be summed up. Finally, the demagnetization factor \eqref{eq:Nx-full-2} takes the form:
\begin{equation} \label{eq:Nx}
\begin{split}
N_x &= \frac{1}{9 \pi  \left(\mu ^{3/2}-1\right)} \Biggl[(\mu -1) (6 \mu +1) \text{K}(\mu )\\
&+(3 \pi -7) \mu ^{3/2}+7 (\mu +1) \text{E}(\mu )-3 \pi -7 \Biggr],\\
\mu & = 1/(1+\varepsilon)^2,
\end{split}
\end{equation}
where $\text{K}(\mu )$, $\text{E}(\mu )$ are the complete elliptic integrals of the first and second kind, respectively \cite{Abramowitz72}. The energy density of the uniform state, normalized by the value $4\pi M_S^2 V$, reads
\begin{equation} \label{eq:Ex}
\mathcal{E}_u = \frac{N_x}{2}.
\end{equation}
The energy $\mathcal{E}_u$ decreases slightly with an aspect ratio, see the solid blue curve in Fig.~\ref{fig:en}.

Note that $N_x\leq 1/3$ and the equality takes place for $\varepsilon=0$, which means that it is easier to magnetize the shell in the cut plane than along the $z$--direction. It is instructive to mention that such a picture differs strongly from the case of extruded hemispheres \cite{Streubel12}. In the latter  case there exist two monodomain states: the onion state is favoured for small enough aspect ratios, when $\varepsilon<\varepsilon_c$ with $\varepsilon_c\approx 1.47$ \cite{Streubel12}. When $\varepsilon > \varepsilon_c$, an easy--axis magnetization distribution is realized. The reason for the uniform easy--axes state is dictated by geometry: the sample is elongated along z--axis, having the shape of cylinder with spherically deformed face surfaces.

\begin{figurehere}
\begin{center}
\includegraphics[width=1.03\columnwidth]{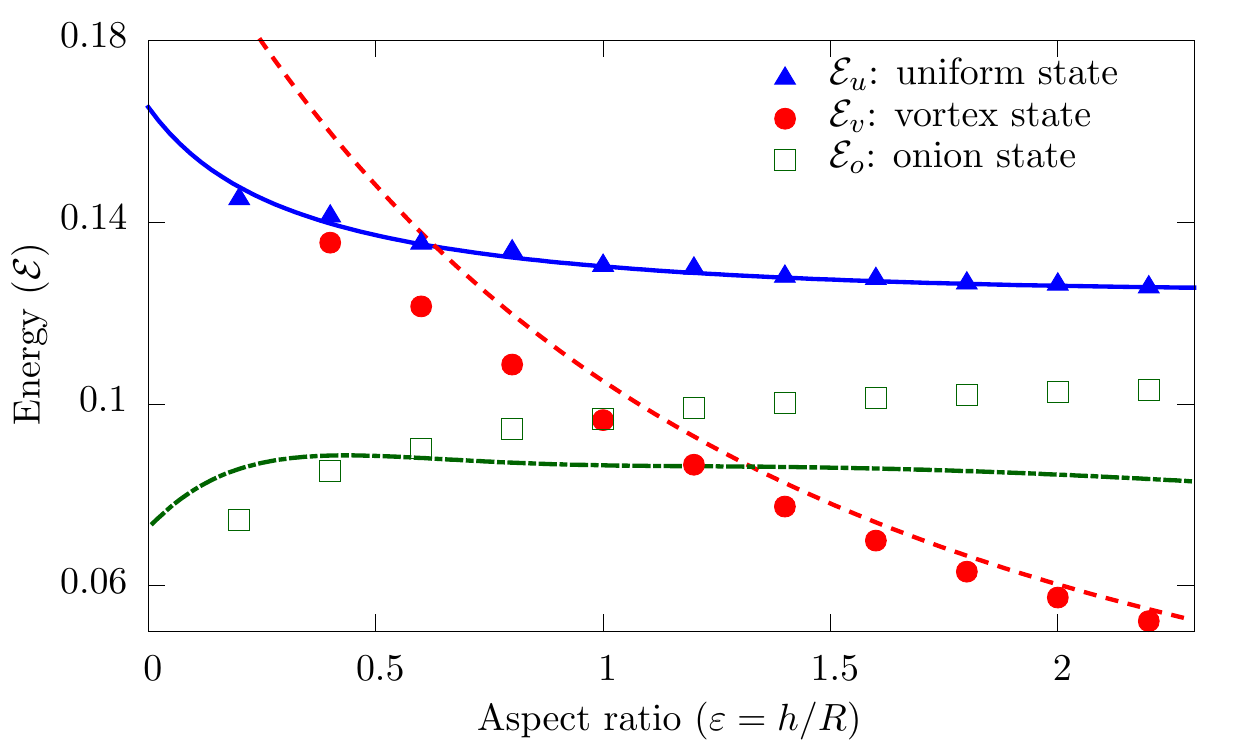}
\end{center}
\caption{Energies of different magnetization structures as functions of the particle aspect ratio. Symbols correspond to the micromagnetic simulations (Py shells with $R=15$ nm and different $h$), lines to analytics: solid blue line describes the dependence \eqref{eq:Ex}, dashed red line corresponds to \eqref{eq:Ev}, and the dash--and--dot green line corresponds to \eqref{q:Enery-onion-min}.}
\label{fig:en}
\end{figurehere}

\section{Vortex state}
\label{sec:vortex}

The monodomain magnetization distribution can form the ground state for relatively small samples. It is well known for flat samples \cite{Hubert98}, that the magnetization curling becomes energetically favorable with the particle size increasing. The reason is the competition between the exchange and stray field. In particular, for the disk shape particle, there appears a vortex state \cite{Hubert98,Guimaraes09}. The static vortex state provides the absence of volume and edge surface magnetostatic charges. The only small stray field comes from face surface charges, which are localized inside the core.

The static vortex configuration can be described similar to the vortex configuration in disks:
\begin{subequations} \label{eq:vortex-ansatz}
\begin{equation} \label{eq:vortex}
\cos\Theta = p f(\rho), \qquad \Phi = q\chi + \mathfrak{C}\pi/2.
\end{equation}
Here we use cylinder coordinates $(\rho,\chi,z)$ for the radius--vector $\vec{r}$, the parameter $q=1$ is the $\pi^1$ vortex charge (vorticity), $p=\pm1$ is the vortex polarity (outward and inward), and $\mathfrak{C}=\pm1$ is the vortex chirality (clockwise or counterclockwise). The function $f(\rho)$ describes a vortex out--of--surface structure, which is localized inside the vortex core with the typical radius $\rho_c$ of the order of the exchange length $\ell$. Here we use the exchange Ansatz by Usov \cite{Usov93}; according to it the vortex out--of--plane structure has the form of Belavin--Polyakov soliton \cite{Belavin75} inside the core, and takes zero value outside:
\begin{equation} \label{eq:Usov}
f(\rho) =
\begin{cases}
\dfrac{\rho_c^2-\rho^2}{\rho_c^2+\rho^2}, &\text{when $\rho<\rho_c$}\\
0, &\text{when $\rho\geq \rho_c$}
\end{cases}
,\quad\rho_c = \beta\ell.
\end{equation}
\end{subequations}
Here $\beta\sim 1$ is the variational parameter.

For the analytical treatment of the model, we make one simplification: we use a local shape--anisotropy instead of the nonlocal stray field, what is acceptably for thin particles \cite{Gioia97,Guslienko04}. In the case of the curved shell one has to consider the easy--surface anisotropy \cite{Kravchuk12a} instead of easy--plane used for the flat magnets. Nevertheless the vortex Ansatz \eqref{eq:vortex-ansatz} has the localized structure out--of--surface, hence for $\rho_c/ R\rightarrow0$ the vortex solution in the easy--surface model has the same structure as in the easy--plane one. Using the angular parametrisation \eqref{eq:m} for the normalised magnetisation $\vec{m}$, we limit ourselves by the following energy functional
\begin{equation} \label{eq:En-ex-n-an}
\begin{split}
\mathcal{E}^{\text{ex}+\text{an}} = \frac{\ell^2}{2V} \int \mathrm{d}^3x \Biggl[& (\vec\nabla\Theta)^2 + \sin^2\Theta(\vec\nabla\Phi)^2\\
& + \frac{\cos^2\Theta}{\ell^2}\Biggr],
\end{split}
\end{equation}
where we assumed that $\ell\ll R$.

Under aforementioned assumptions we derive the vortex state energy \eqref{eq:En-ex-n-an} using the vortex Ansatz \eqref{eq:vortex-ansatz}:
\begin{equation} \label{eq:En-vor-2}
\begin{split}
\mathcal{E}_v &= \frac{\pi \ell^2 h}{V} \int\limits_0^{\rho_c} \rho \mathrm{d}\rho \left( \frac{1-f^2}{\rho^2} + \frac{{f'}^2}{1-f^2} + \frac{f^2}{\ell^2} \right)\\
& + \frac{\pi \ell^2}{V}\Biggl( \int\limits_{\rho_c}^R \frac{\mathrm{d}\rho}{\rho} \int\limits_{\sqrt{R^2-\rho^2}}^{\sqrt{(R+h)^2-\rho^2}} \mathrm{d}z\\
& + \int\limits_{R}^{R+h} \frac{\mathrm{d}\rho}{\rho} \int\limits_0^{\sqrt{(R+h)^2-\rho^2}} \mathrm{d}z \Biggr),
\end{split}
\end{equation}
where the integration over $\rho$ was splitted into three domains: $0<\rho<\rho_c$, $\rho_c<\rho<R$ and $R<\rho<R+h$; we also assumed that the shape of the first domain is cylindrical one. To simplify the further analysis we suppose that exchange length $\ell$ is much smaller than the system size $R$. In this case the energy \eqref{eq:En-vor-2} can be written as follows
\begin{equation} \label{eq:En-vor-3}
\begin{split}
\mathcal{E}_v \approx &\frac{\lambda ^2}{2\varepsilon  (1 + \varepsilon + \varepsilon^2/3)} \Biggl[(1+\varepsilon)\ln (1+\varepsilon) \\
&+ \varepsilon\left(1+\ln\frac{2}{\beta\lambda} + \frac{\beta^2}{2\beta_0^2} \right)\Biggr], \\
& \beta_0 = \frac{1}{\sqrt{3-4\ln2}}\approx 2.097.
\end{split}
\end{equation}
Here $\lambda = \ell/R\ll1$ is a reduced exchange length. By minimizing the energy \eqref{eq:En-vor-3} with respect to $\beta$, one can find $\beta =\beta_0$. Finally, the energy of the vortex state shell, which corresponds to the optimized value of $\beta$, reads:
\begin{equation} \label{eq:Ev}
\begin{split}
\mathcal{E}_v \approx &\frac{\lambda ^2 \left[(1+\varepsilon)\ln (1+\varepsilon) + \varepsilon\left(3/2 - \ln\lambda \right)\right] }{2\varepsilon  (1 + \varepsilon + \varepsilon^2/3)}.
\end{split}
\end{equation}
The dependence \eqref{eq:Ev} is well confirmed by simulations, see the dashed red curve in Fig.~\ref{fig:en}.

\section{Onion state}
\label{sec:onion}

One more nontrivial magnetization configuration, which is realized in magnetic cap, is the onion state. Such a configuration is well known for the ring geometry, where the onion state appears as a high remanence state. Onion states for the ring geometry were intensively studied  during the last decade experimentally \cite{Klaui01,Lopez01,Klaui03a,Zhang10} and theoretically \cite{Landeros06}. The magnetization configuration in the onion state ring in each half of the ring has an opposite sense  of circulation, forming two edge solitons (boojums). The onion configuration is also known for the spherical shells \cite{Kong08a,Kravchuk12a}. Very recently we observed the onion state in extruded hemispheres \cite{Streubel12}.

In order to describe the onion state in the hemispherical caps, we propose the following \emph{one--parameter Ansatz}
\begin{equation} \label{eq:onion-Ansatz}
\begin{split}
\vec{m} &= \left(-\sin\theta, \cos\theta\cos\phi, \cos\theta\sin\phi \right),\\
\cos\theta &=\frac{x}{r}, \qquad \tan\phi = \frac{z+a\,r}{y}
\end{split}
\end{equation}
with $\vec{r}=\left(x,y,z\right)$ being the radius--vector. Here $a>0$ is the parameter of the model, which describes the position of singularities: the case $a=0$ corresponds to meridian--like magnetization distribution, for $a>0$ the position of the singularities goes down in $z$--direction, for the limit case $a\gg1$ the magnetization distribution in the ring  surface ($z=0$) becomes almost parallel, while it is still quasi--tangential to the sphere in the outer and inner surfaces. The typical onion configuration, described by Ansatz \eqref{eq:onion-Ansatz} is presented in Fig~\ref{fig:onion}.

\begin{figurehere}
\begin{center}
\begin{tikzpicture}[scale=1]
{%
\node[right,xshift=0,yshift=7] at (0,3.5) {\includegraphics[width=0.4\columnwidth]{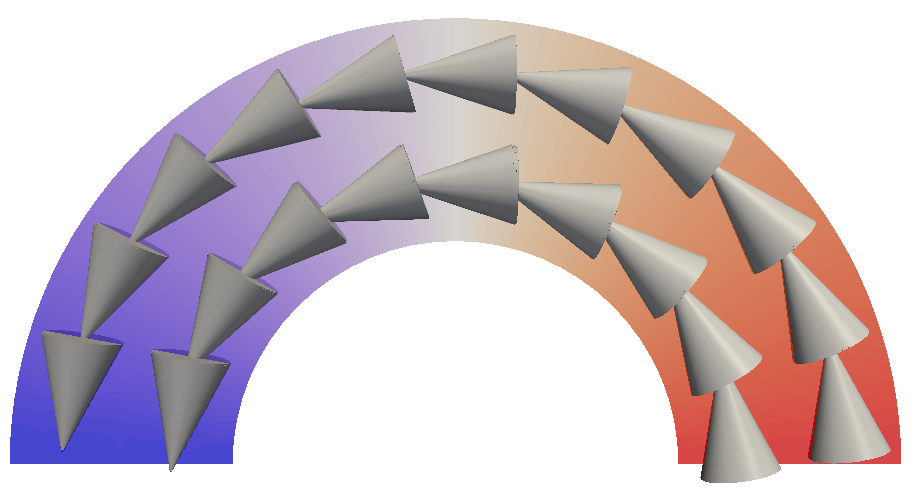}};
\node[right,xshift=0,yshift=-10] at (0,1.75) {\includegraphics[width=0.4\columnwidth]{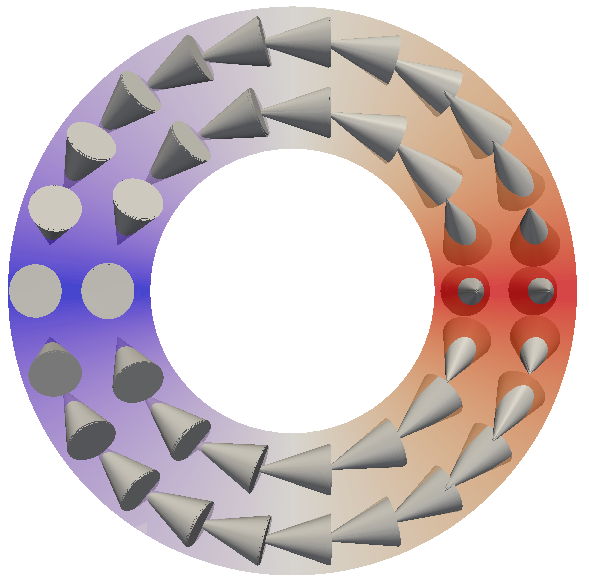}};
\node[right,xshift=0,yshift=0] at (1,-0.65) {(a) $a=0.5$.};
} %
{%
\node[right,xshift=100,yshift=7] at (0,3.5) {\includegraphics[width=0.4\columnwidth]{OnionAnzatzSide.png}};
\node[right,xshift=100,yshift=-10] at (0,1.75) {\includegraphics[width=0.4\columnwidth]{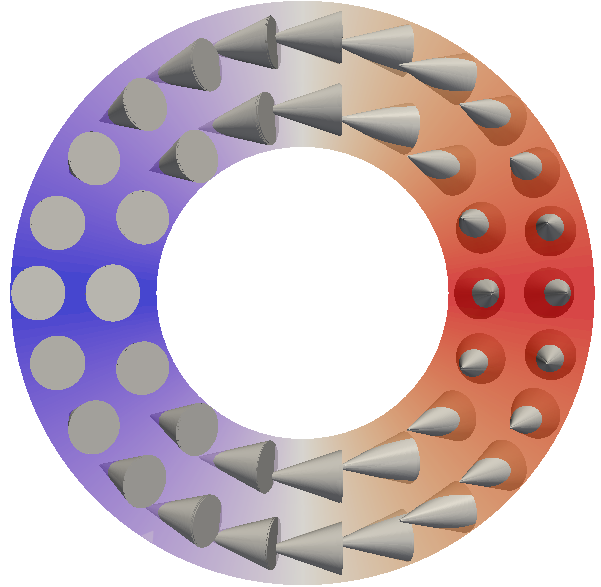}};
\node[right,xshift=100,yshift=0] at (1,-0.65) {(b) $a=5.0$.};
\node[right,xshift=5,yshift=0] at (7.1,2) {\includegraphics[width=0.3\columnwidth,angle=90]{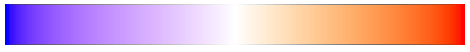}};
\node[right,xshift=5,yshift=0] at (7.4,0.8) {\scriptsize $-1$};
\node[right,xshift=5,yshift=0] at (7.5,3.3) {\scriptsize $1$};
\node[right,xshift=5,yshift=0] at (7.4,2) {\small $m_z$};
} %
%
\end{tikzpicture}
\end{center}
\caption{Typical onion configuration from the Ansatz \eqref{eq:onion-Ansatz}.}
\label{fig:onion}
\end{figurehere}

The energy functional consists of the exchange term and the magnetostatic one. Similar to the previous section we replace the magnetostatic energy of the spherical surfaces by its local counterpart, the easy--surface anisotropy. This contribution is described by the functional $\mathcal{E}^{\text{ex}+\text{an}}$, see \eqref{eq:En-ex-n-an}. When we described the vortex state it was sufficient to limit ourselves by such approach since the vortex magnetization distribution does not produce neither  volume charges nor surface charges on the ring surface. It is not the case for the onion state. However, a detailed numerical analysis shows that both volume magnetic charge contribution and the contribution due to the interaction between the spherical surface charges and ring surface charges are negligibly small  as compared with $\mathcal{E}^{\text{ex}+\text{an}}$.

Analysis shows that the essential role in the energy balance plays the magnetostatic energy of the ring surface itself
\begin{align} \label{eq:E-ring-1}
&\mathscr{E}^{\text{ring}} = \frac{1}{8\pi V} \!\!\!\int\limits_{S^{\text{ring}}} \!\!\! \mathrm{d}S^{\text{ring}}\!\!\! \int\limits_{{S'}^{\text{ring}}} \!\!\!\mathrm{d}{S'}^{\text{ring}}\, \frac{m_z m'_z }{\left| \vec{r} - \vec{r}'\right|}\\
&=A_\varepsilon \!\!\int\limits_0^1\!\! \mathrm{d}\xi\!\!\int\limits_0^1 \!\!\mathrm{d}\xi'\!\! \int\limits_0^{2\pi}\!\! \mathrm{d}\chi \!\!\int\limits_0^{2\pi} \!\! \mathrm{d}\chi' \frac{\sigma(\chi)\sigma(\chi')\left(1+\varepsilon\xi \right) \left(1+\varepsilon\xi' \right)}{R(\xi,\xi',\chi-\chi')}.
\nonumber
\end{align}
Here the source of the energy is the ring surface charges
\begin{equation} \label{eq:sigma-ring}
\sigma(\chi) = \frac{a\cos\chi}{\sqrt{a^2+\sin^2\chi}},
\end{equation}
the other parameters are as follows:
\begin{equation} \label{eq:A-ring}
\begin{split}
A_\varepsilon &= \frac{\varepsilon}{16 \pi^2 \left(1+\varepsilon+\varepsilon^2/3 \right)},\\
R(\xi,\xi',\alpha) &= \Bigl[(1+\varepsilon\xi)^2+ (1+\varepsilon\xi')^2\\
&- 2(1+\varepsilon\xi)(1+\varepsilon\xi')\cos(\alpha)\Bigr]^{1/2}.
\end{split}
\end{equation}

In the same way we incorporate the Ansatz \eqref{eq:onion-Ansatz} into the energy functional \eqref{eq:En-ex-n-an}. As a result the energy of the onion state reads:
\begin{subequations} \label{q:Enery-onion-min}
\begin{equation} \label{eq:Energy-onion}
\mathscr{E}_o = \mathcal{E}^{\text{ex}+\text{an}} + \mathscr{E}^{\text{ring}}.
\end{equation}
Using numerical integration we computed the energy of the onion state, which corresponds to the optimal parameters $a_o$ from the condition
\begin{equation} \label{eq:ao}
\frac{\partial \mathscr{E}_o}{\partial a}\Bigr|_{a=a_o} = 0.
\end{equation}
\end{subequations}
The energy of the onion state \eqref{q:Enery-onion-min} linearly increases with the aspect ratio for small $\varepsilon$ (due to the dependence $A_\varepsilon\propto \varepsilon$) and rapidly goes to the saturation value for higher $\varepsilon$ mainly, see the dash--and--dot green curve in Fig.~\ref{fig:en}.

One has to note that the energy of the onion state is always smaller than the homogeneous one, hence it can form the ground state in hemispherical caps.

\section{Phase diagram}
\label{sec:PhD}

\begin{figure*}
\begin{center}
\begin{tikzpicture}[scale=1]
\node[right] at (0,0) {\includegraphics[width=2\columnwidth]{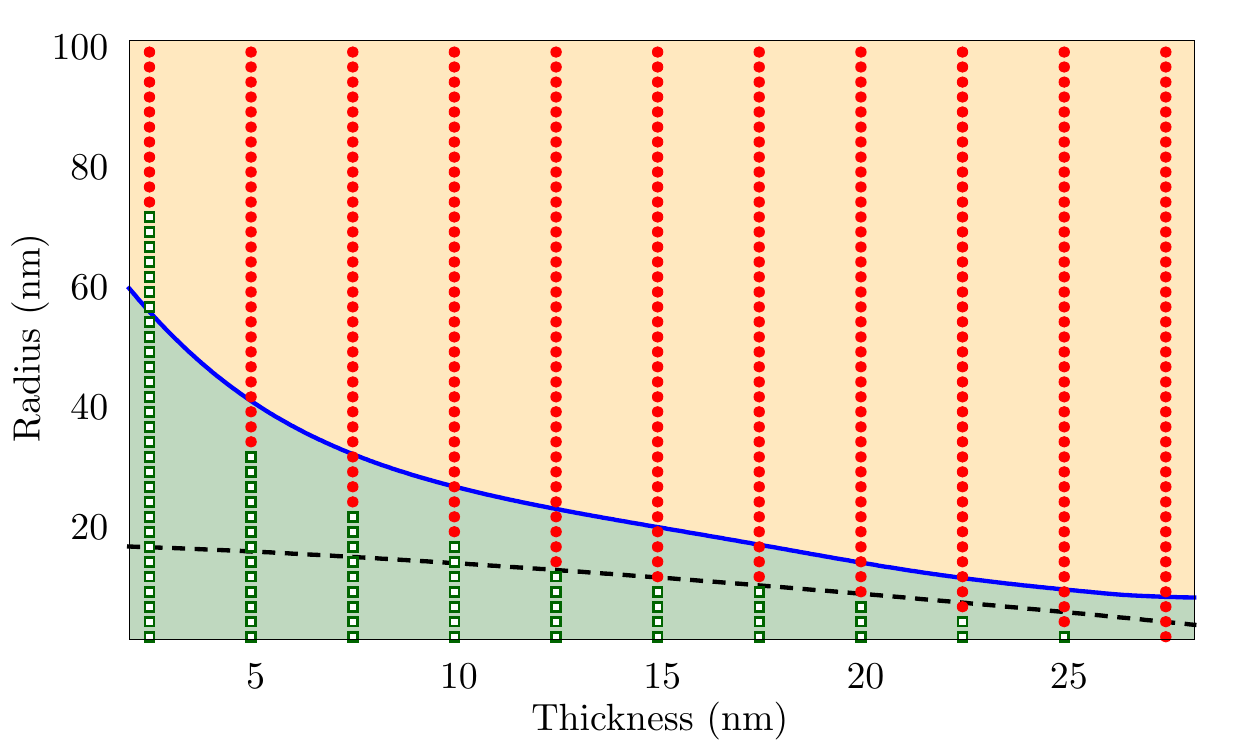}};
\filldraw[very thin, fill=blue!15!white,opacity=0.95,rounded corners=10pt,xshift=0,yshift=10] (3,-1) --  (7.5,-1) -- (6.35,-4) -- (8,-1) --  (9,-1) -- (9,2.5) -- (3,2.5) -- cycle;
\node[right,xshift=0,yshift=10] at (3,1.5) {\includegraphics[width=0.3\columnwidth]{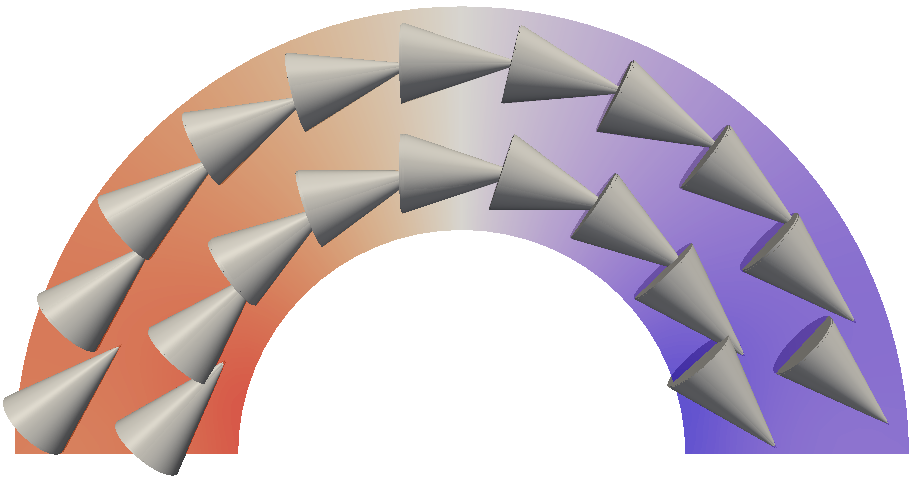}};
\node[right,xshift=0,yshift=10] at (6,1) {\includegraphics[width=0.3\columnwidth]{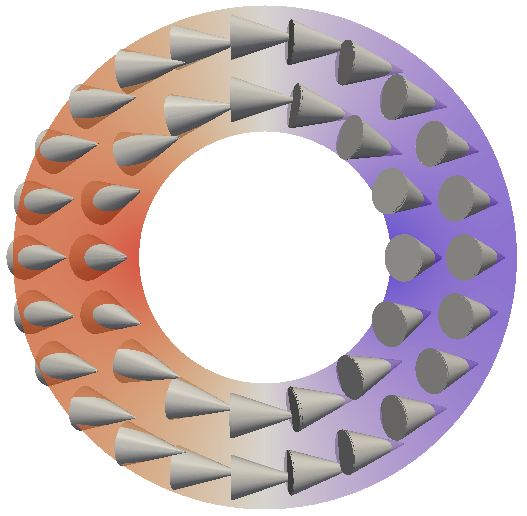}};
\node[right,xshift=0,yshift=10] at (3.4,0.1) {\includegraphics[width=0.2\columnwidth]{scale_m}};
\node[right,xshift=0,yshift=10] at (3,0.1) {\scriptsize $-1$};
\node[right,xshift=0,yshift=10] at (5.15,0.1) {\scriptsize $1$};
\node[right,xshift=0,yshift=10] at (4,0.33) {\small $m_z$};
\node[right,xshift=20,yshift=10] at (3.5,-0.5) {\Large \bf Onion state};
\filldraw[very thin, fill=blue!15!white,opacity=0.95, rounded corners=10pt,xshift=200,yshift=40] (3,-1) -- (7,-1) -- (6.3,-3) -- (7.5,-1) --  (9,-1) -- (9,2.5) -- (3,2.5) -- cycle;
\node[right,xshift=200,yshift=40] at (3,1.5) {\includegraphics[width=0.3\columnwidth]{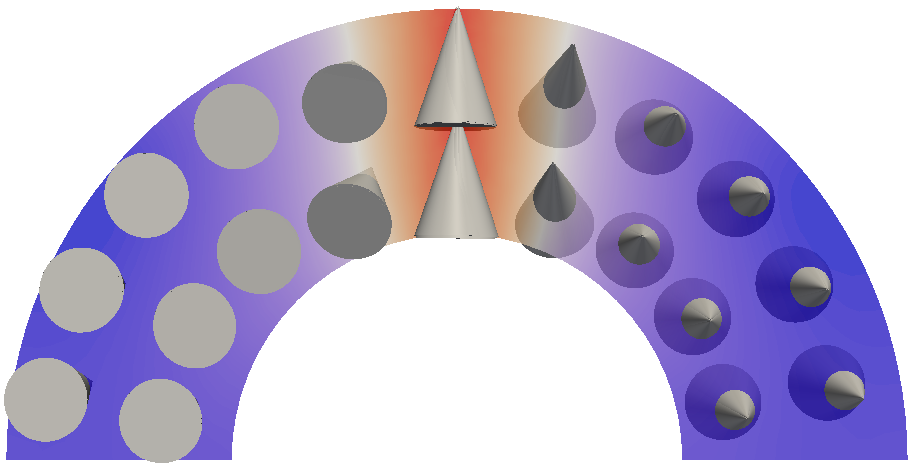}};
\node[right,xshift=200,yshift=40] at (6,1) {\includegraphics[width=0.3\columnwidth]{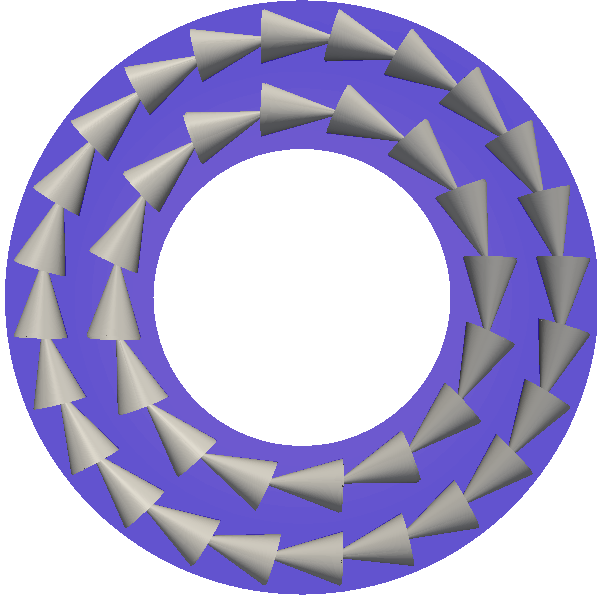}};
\node[right] at (10.45,1.45) {\includegraphics[width=0.2\columnwidth]{scale_m}};
\node[right] at (10.3,1.45) {\scriptsize $0$};
\node[right] at (12.2,1.45) {\scriptsize $1$};
\node[right] at (11.2,1.65) {\small $m_z$};
\node[right,xshift=220,yshift=40] at (3.5,-0.5) {\Large \bf Vortex state};
%
\end{tikzpicture}
\end{center}

\caption{Phase diagrams of equilibrium magnetization structures in the hemispherical shell. Symbols correspond to simulation data for Py nanoshellps: open green squares to the monodomain state and red circles to  the vortex one. The solid line corresponds to theoretically calculated border between the onion state and the vortex one as the numerical solution of \eqref{eq:onion-vortex}. The dashed line is calculated from the energy balance between the uniform state and the vortex one, which is determined bu \eqref{eq:uniform-vortex}. Insets show magnetization distribution for onion and vortex state by micromagnetic simulations: two cross-sections correspond to cut surfaces from Fig.~\ref{fig:mag-full}.}
\label{fig:phd}
\end{figure*}

Let us summarize results on the equilibrium magnetization distribution. By comparing energies of different states, one can calculate the energetically preferable states for different samples sizes. The computed phase diagram is presented in Fig.~\ref{fig:phd}. The general properties of the phase diagram are as follows. The ground state of the very thin hemispherical shell is the monodomain (onion) state. By increasing the shell thickness one can switch to the vortex state.

The boundary between two phases can be derived as follows. Supposing that the monodomain state is the onion one, described by \eqref{q:Enery-onion-min}, and the vortex state by \eqref{eq:Ev}, one can compute the border, which separates different phases, from the condition:
\begin{equation} \label{eq:onion-vortex}
\mathcal{E}_o =\mathcal{E}_v.
\end{equation}
The numerical solution of this equation is plotted in Fig.~\ref{fig:phd} by the solid (blue) line. Below this border the onion state is realized, in the upper region the vortex state appears. One has to note that our theoretical treatments are based on Ansatz functions and gives only approximate description of the problem. It is also instructive to compare energies of the uniform state and the vortex one:
\begin{equation} \label{eq:uniform-vortex}
\mathcal{E}_u =\mathcal{E}_v.
\end{equation}
Since the energy of the uniform state is always larger than the onion one, the condition \eqref{eq:uniform-vortex} gives the minorant estimation for the boundary between different phases.

\section{Micromagnetic simulations}

In order to verify analytical predictions we performed micromagnetic simulations of energy minimization procedure using \textsf{MAGPAR} simulator \cite{magpar,Scholz03a} with Permalloy parameters: exchange constant $A = 1.3\times 10^{-11}$ J/m, saturation magnetization $M_s = 8.6\times 10^5$ A/m, and on--site anisotropy was neglected. The mesh size was varied from 0.5 to 4 nm depending on the sample size. For each cap we make two simulations with different initial states, namely the rough vortex state and the uniform state along $x$-axis.

The energy of the uniform initial state is shown by blue triangles in Fig.~\ref{fig:en}. One can see that our analytical approach \eqref{eq:Nx} agrees very well with the simulated data. Numerically we determined the remanent state using the energy minimization procedure. We observed in simulations that the magnetization was relaxed to the onion state, where the magnetization distribution is close to uniform in the top view, while it is almost tangential in the side view [Fig.~\ref{fig:phd}(left inset)], cf. Fig.~\ref{fig:onion}. The energy of the onion state is almost constant for not very small aspect ratios. One can see from the Fig.~\ref{fig:en} that our theoretically calculated curves underestimate the energy of the vortex state for $\varepsilon>0.5$. The reason is that we do not take into account the contributions to the magnetostatic energy by the volume charges, by the interaction between spherical surface charges and the ring surface charge, and by the interaction of volume and surface charges. Nevertheless we see that out theoretically calculated energy gives rather good approximation for the simulated data.

In order to study the vortex state we used the vortex magnetization state with polarity $p=1$ and chirality $\mathfrak{C}=1$ [Fig.~\ref{fig:phd}(right inset)] as initial one and minimize the energy for this case. Red circles in Fig.~\ref{fig:en} correspond to the energy of the obtained remanent vortex state. By comparing energies of vortex and onion states one can see that the vortex state is energetically preferable one when $\varepsilon\gtrsim1$ for the caps with $R=15$ nm, see Fig.~\ref{fig:en}.

To check our theoretically calculated phase diagram, we use 3D micromagnetic simulations, as described above. The comparison of our theory with simulation data shows that the analytically calculated boundary \eqref{eq:onion-vortex} is located upper than the simulations data. The reason is that we underestimate the energy of the onion state, see above, that is why our curve majorizes the correct boundary curve for not very small thicknesses, see the blue solid curve in Fig.~\ref{fig:phd}. We also plotted the curve, which results from the energy balance between the uniform state and the vortex one, see \eqref{eq:uniform-vortex}, which minorizes the correct boundary.

One has to mention one more factor, which causes deviations between analytical results and simulations data: we limit ourselves in description of the vortex and onion states by large sample radii ($R\gg \ell$), see \eqref{eq:En-ex-n-an}. That is why our analytical description of the vortex state \eqref{eq:En-vor-3} and the onion one \eqref{q:Enery-onion-min} deviates from numerics for samples with small inner radius.

\section{Conclusions}

In conclusion, we presented a detailed study of the ground state of hemispherical magnetic nanoshells, including monodomain (onion) state and the vortex one. We proposed a simple analytical description of such states and studied analytically the phase transition between different ground states. As opposed to disks \cite{Metlov02} and extruded hemispheres \cite{Streubel12}, the phase diagram does not contain the uniform easy--axis state. All results are confirmed by our direct 3D \textsf{MAGPAR} micromagnetic simulations.

\nonumsection{Acknowledgments} \noindent  A. Vidil (Kiev University) is acknowledged for participation at the early stage of the project. All simulations results presented in the work were obtained using the computing cluster of Kiev University \cite{unicc}.

%

\begin{thebibliography}{25}
\expandafter\ifx\csname natexlab\endcsname\relax\def\natexlab#1{#1}\fi
\expandafter\ifx\csname bibnamefont\endcsname\relax
  \def\bibnamefont#1{#1}\fi
\expandafter\ifx\csname bibfnamefont\endcsname\relax
  \def\bibfnamefont#1{#1}\fi
\expandafter\ifx\csname citenamefont\endcsname\relax
  \def\citenamefont#1{#1}\fi
\expandafter\ifx\csname url\endcsname\relax
  \def\url#1{\texttt{#1}}\fi
\expandafter\ifx\csname urlprefix\endcsname\relax\def\urlprefix{URL }\fi
\providecommand{\bibinfo}[2]{#2}
\providecommand{\eprint}[2][]{\url{#2}}

\bibitem[{\citenamefont{Guimar$\tilde{\mathrm{a}}$es}(2009)}]{Guimaraes09}
\bibinfo{author}{\bibfnamefont{A.~P.}
  \bibnamefont{Guimar$\tilde{\mathrm{a}}$es}}, \emph{\bibinfo{title}{Principles
  of Nanomagnetism}}, NanoScience and Technology
  (\bibinfo{publisher}{Springer-Verlag Berlin Heidelberg},
  \bibinfo{year}{2009}), ISBN \bibinfo{isbn}{978-3-642-01481-9}.

\bibitem[{\citenamefont{Bohlens et~al.}(2008)\citenamefont{Bohlens, Kr\"{u}ger,
  Drews, Bolte, Meier, and Pfannkuche}}]{Bohlens08}
\bibinfo{author}{\bibfnamefont{S.}~\bibnamefont{Bohlens}},
  \bibinfo{author}{\bibfnamefont{B.}~\bibnamefont{Kr\"{u}ger}},
  \bibinfo{author}{\bibfnamefont{A.}~\bibnamefont{Drews}},
  \bibinfo{author}{\bibfnamefont{M.}~\bibnamefont{Bolte}},
  \bibinfo{author}{\bibfnamefont{G.}~\bibnamefont{Meier}}, \bibnamefont{and}
  \bibinfo{author}{\bibfnamefont{D.}~\bibnamefont{Pfannkuche}},
  \bibinfo{journal}{Appl. Phys. Lett.} \textbf{\bibinfo{volume}{93}},
  \bibinfo{eid}{142508} (pages~\bibinfo{numpages}{3}) (\bibinfo{year}{2008}),
  \urlprefix\url{http://link.aip.org/link/?APL/93/142508/1}.

\bibitem[{\citenamefont{Pigeau et~al.}(2010)\citenamefont{Pigeau, de~Loubens,
  Klein, Riegler, Lochner, Schmidt, Molenkamp, Tiberkevich, and
  Slavin}}]{Pigeau10}
\bibinfo{author}{\bibfnamefont{B.}~\bibnamefont{Pigeau}},
  \bibinfo{author}{\bibfnamefont{G.}~\bibnamefont{de~Loubens}},
  \bibinfo{author}{\bibfnamefont{O.}~\bibnamefont{Klein}},
  \bibinfo{author}{\bibfnamefont{A.}~\bibnamefont{Riegler}},
  \bibinfo{author}{\bibfnamefont{F.}~\bibnamefont{Lochner}},
  \bibinfo{author}{\bibfnamefont{G.}~\bibnamefont{Schmidt}},
  \bibinfo{author}{\bibfnamefont{L.~W.} \bibnamefont{Molenkamp}},
  \bibinfo{author}{\bibfnamefont{V.~S.} \bibnamefont{Tiberkevich}},
  \bibnamefont{and} \bibinfo{author}{\bibfnamefont{A.~N.}
  \bibnamefont{Slavin}}, \bibinfo{journal}{Appl. Phys. Lett.}
  \textbf{\bibinfo{volume}{96}}, \bibinfo{pages}{132506}
  (\bibinfo{year}{2010}), ISSN \bibinfo{issn}{00036951},
  \urlprefix\url{http://dx.doi.org/10.1063/1.3373833}.

\bibitem[{\citenamefont{Yu et~al.}(2011)\citenamefont{Yu, Jung, Lee, Fischer,
  and Kim}}]{Yu11a}
\bibinfo{author}{\bibfnamefont{Y.-S.} \bibnamefont{Yu}},
  \bibinfo{author}{\bibfnamefont{H.}~\bibnamefont{Jung}},
  \bibinfo{author}{\bibfnamefont{K.-S.} \bibnamefont{Lee}},
  \bibinfo{author}{\bibfnamefont{P.}~\bibnamefont{Fischer}}, \bibnamefont{and}
  \bibinfo{author}{\bibfnamefont{S.-K.} \bibnamefont{Kim}},
  \bibinfo{journal}{Appl. Phys. Lett.} \textbf{\bibinfo{volume}{98}},
  \bibinfo{eid}{052507} (pages~\bibinfo{numpages}{3}) (\bibinfo{year}{2011}),
  \urlprefix\url{http://link.aip.org/link/?APL/98/052507/1}.

\bibitem[{\citenamefont{Hertel}(2013)}]{Hertel13}
\bibinfo{author}{\bibfnamefont{R.}~\bibnamefont{Hertel}}, \bibinfo{journal}{Nat
  Nano} \textbf{\bibinfo{volume}{8}}, \bibinfo{pages}{318}
  (\bibinfo{year}{2013}), ISSN \bibinfo{issn}{1748-3387},
  \urlprefix\url{http://dx.doi.org/10.1038/nnano.2013.81}.

\bibitem[{\citenamefont{Uhl\'{\i}\u{r}
  et~al.}(2013)\citenamefont{Uhl\'{\i}\u{r}, Urb\'{a}nek, Hlad\'{\i}k, Spousta,
  Im, Fischer, Eibagi, Kan, and \u{S}ikola}}]{Uhlir13}
\bibinfo{author}{\bibfnamefont{V.}~\bibnamefont{Uhl\'{\i}\u{r}}},
  \bibinfo{author}{\bibfnamefont{M.}~\bibnamefont{Urb\'{a}nek}},
  \bibinfo{author}{\bibfnamefont{L.}~\bibnamefont{Hlad\'{\i}k}},
  \bibinfo{author}{\bibfnamefont{J.}~\bibnamefont{Spousta}},
  \bibinfo{author}{\bibfnamefont{M.-Y.} \bibnamefont{Im}},
  \bibinfo{author}{\bibfnamefont{P.}~\bibnamefont{Fischer}},
  \bibinfo{author}{\bibfnamefont{N.}~\bibnamefont{Eibagi}},
  \bibinfo{author}{\bibfnamefont{E.~E.} \bibnamefont{Kan},
  \bibfnamefont{J.~J.~Fullerton}}, \bibnamefont{and}
  \bibinfo{author}{\bibfnamefont{T.}~\bibnamefont{\u{S}ikola}},
  \bibinfo{journal}{Nat Nano} \textbf{\bibinfo{volume}{8}},
  \bibinfo{pages}{341} (\bibinfo{year}{2013}), ISSN \bibinfo{issn}{1748-3387},
  \urlprefix\url{http://dx.doi.org/10.1038/nnano.2013.66}.

\bibitem[{\citenamefont{Hubert and Sch{\" a}fer}(1998)}]{Hubert98}
\bibinfo{author}{\bibfnamefont{A.}~\bibnamefont{Hubert}} \bibnamefont{and}
  \bibinfo{author}{\bibfnamefont{R.}~\bibnamefont{Sch{\" a}fer}},
  \emph{\bibinfo{title}{Magnetic domains: the analysis of magnetic
  microstructures}} (\bibinfo{publisher}{Springer--Verlag},
  \bibinfo{address}{Berlin}, \bibinfo{year}{1998}).

\bibitem[{\citenamefont{Kravchuk et~al.}(2012)\citenamefont{Kravchuk, Sheka,
  Streubel, Makarov, Schmidt, and Gaididei}}]{Kravchuk12a}
\bibinfo{author}{\bibfnamefont{V.~P.} \bibnamefont{Kravchuk}},
  \bibinfo{author}{\bibfnamefont{D.~D.} \bibnamefont{Sheka}},
  \bibinfo{author}{\bibfnamefont{R.}~\bibnamefont{Streubel}},
  \bibinfo{author}{\bibfnamefont{D.}~\bibnamefont{Makarov}},
  \bibinfo{author}{\bibfnamefont{O.~G.} \bibnamefont{Schmidt}},
  \bibnamefont{and} \bibinfo{author}{\bibfnamefont{Y.}~\bibnamefont{Gaididei}},
  \bibinfo{journal}{Phys. Rev. B} \textbf{\bibinfo{volume}{85}},
  \bibinfo{pages}{144433} (\bibinfo{year}{2012}),
  \urlprefix\url{http://link.aps.org/doi/10.1103/PhysRevB.85.144433}.

\bibitem[{\citenamefont{Streubel
  et~al.}(2012{\natexlab{a}})\citenamefont{Streubel, Kravchuk, Sheka, Makarov,
  Kronast, Schmidt, and Gaididei}}]{Streubel12}
\bibinfo{author}{\bibfnamefont{R.}~\bibnamefont{Streubel}},
  \bibinfo{author}{\bibfnamefont{V.~P.} \bibnamefont{Kravchuk}},
  \bibinfo{author}{\bibfnamefont{D.~D.} \bibnamefont{Sheka}},
  \bibinfo{author}{\bibfnamefont{D.}~\bibnamefont{Makarov}},
  \bibinfo{author}{\bibfnamefont{F.}~\bibnamefont{Kronast}},
  \bibinfo{author}{\bibfnamefont{O.~G.} \bibnamefont{Schmidt}},
  \bibnamefont{and} \bibinfo{author}{\bibfnamefont{Y.}~\bibnamefont{Gaididei}},
  \bibinfo{journal}{Appl. Phys. Lett.} \textbf{\bibinfo{volume}{101}},
  \bibinfo{eid}{132419} (pages~\bibinfo{numpages}{3})
  (\bibinfo{year}{2012}{\natexlab{a}}),
  \urlprefix\url{http://link.aip.org/link/?APL/101/132419/1}.

\bibitem[{\citenamefont{Streubel
  et~al.}(2012{\natexlab{b}})\citenamefont{Streubel, Makarov, Kronast,
  Kravchuk, Albrecht, and Schmidt}}]{Streubel12a}
\bibinfo{author}{\bibfnamefont{R.}~\bibnamefont{Streubel}},
  \bibinfo{author}{\bibfnamefont{D.}~\bibnamefont{Makarov}},
  \bibinfo{author}{\bibfnamefont{F.}~\bibnamefont{Kronast}},
  \bibinfo{author}{\bibfnamefont{V.}~\bibnamefont{Kravchuk}},
  \bibinfo{author}{\bibfnamefont{M.}~\bibnamefont{Albrecht}}, \bibnamefont{and}
  \bibinfo{author}{\bibfnamefont{O.~G.} \bibnamefont{Schmidt}},
  \bibinfo{journal}{Phys. Rev. B} \textbf{\bibinfo{volume}{85}},
  \bibinfo{pages}{174429} (\bibinfo{year}{2012}{\natexlab{b}}),
  \urlprefix\url{http://link.aps.org/doi/10.1103/PhysRevB.85.174429}.

\bibitem[{\citenamefont{Abramowitz and Stegun}(1972)}]{Abramowitz72}
\bibinfo{author}{\bibfnamefont{M.}~\bibnamefont{Abramowitz}} \bibnamefont{and}
  \bibinfo{author}{\bibfnamefont{I.~A.} \bibnamefont{Stegun}},
  \emph{\bibinfo{title}{Handbook of mathematical functions with formulas,
  graphs, and mathematical tables}} (\bibinfo{publisher}{Dover},
  \bibinfo{address}{New York}, \bibinfo{year}{1972}), \bibinfo{edition}{ninth
  {D}over printing, tenth {GPO} printing} ed., ISBN
  \bibinfo{isbn}{0-486-61272-4}.

\bibitem[{\citenamefont{Usov and Peschany}(1993)}]{Usov93}
\bibinfo{author}{\bibfnamefont{N.~A.} \bibnamefont{Usov}} \bibnamefont{and}
  \bibinfo{author}{\bibfnamefont{S.~E.} \bibnamefont{Peschany}},
  \bibinfo{journal}{J.~Magn. Magn. Mater.} \textbf{\bibinfo{volume}{118}},
  \bibinfo{pages}{L290} (\bibinfo{year}{1993}),
  \urlprefix\url{http://www.sciencedirect.com/science/article/B6TJJ-46FHH1R-C8/2/24f0bbd22588428bbea2c69d514bc837}.

\bibitem[{\citenamefont{Belavin and Polyakov}(1975)}]{Belavin75}
\bibinfo{author}{\bibfnamefont{A.~A.} \bibnamefont{Belavin}} \bibnamefont{and}
  \bibinfo{author}{\bibfnamefont{A.~M.} \bibnamefont{Polyakov}},
  \bibinfo{journal}{JETP Lett.} \textbf{\bibinfo{volume}{22}},
  \bibinfo{pages}{245} (\bibinfo{year}{1975}).

\bibitem[{\citenamefont{Gioia and James}(1997)}]{Gioia97}
\bibinfo{author}{\bibfnamefont{G.}~\bibnamefont{Gioia}} \bibnamefont{and}
  \bibinfo{author}{\bibfnamefont{R.~D.} \bibnamefont{James}},
  \bibinfo{journal}{Proc. R. Soc. Lond. A} \textbf{\bibinfo{volume}{453}},
  \bibinfo{pages}{213} (\bibinfo{year}{1997}),
  \urlprefix\url{http://www.journals.royalsoc.ac.uk/openurl.asp?genre=article&id=doi:10.1098/rspa.1997.0013}.

\bibitem[{\citenamefont{Guslienko and Novosad}(2004)}]{Guslienko04}
\bibinfo{author}{\bibfnamefont{K.~Y.} \bibnamefont{Guslienko}}
  \bibnamefont{and} \bibinfo{author}{\bibfnamefont{V.}~\bibnamefont{Novosad}},
  \bibinfo{journal}{J.~Appl. Phys.} \textbf{\bibinfo{volume}{96}},
  \bibinfo{pages}{4451} (\bibinfo{year}{2004}),
  \urlprefix\url{http://link.aip.org/link/?JAP/96/4451/1}.

\bibitem[{\citenamefont{Kl\"{a}ui et~al.}(2001)\citenamefont{Kl\"{a}ui,
  Rothman, Lopez-Diaz, Vaz, Bland, and Cui}}]{Klaui01}
\bibinfo{author}{\bibfnamefont{M.}~\bibnamefont{Kl\"{a}ui}},
  \bibinfo{author}{\bibfnamefont{J.}~\bibnamefont{Rothman}},
  \bibinfo{author}{\bibfnamefont{L.}~\bibnamefont{Lopez-Diaz}},
  \bibinfo{author}{\bibfnamefont{C.~A.~F.} \bibnamefont{Vaz}},
  \bibinfo{author}{\bibfnamefont{J.~A.~C.} \bibnamefont{Bland}},
  \bibnamefont{and} \bibinfo{author}{\bibfnamefont{Z.}~\bibnamefont{Cui}},
  \bibinfo{journal}{Applied Physics Letters} \textbf{\bibinfo{volume}{78}},
  \bibinfo{pages}{3268} (\bibinfo{year}{2001}),
  \urlprefix\url{http://link.aip.org/link/?APL/78/3268/1}.

\bibitem[{\citenamefont{Lopez-Diaz et~al.}(2001)\citenamefont{Lopez-Diaz,
  Rothman, Kl\"{a}ui, and Bland}}]{Lopez01}
\bibinfo{author}{\bibfnamefont{L.}~\bibnamefont{Lopez-Diaz}},
  \bibinfo{author}{\bibfnamefont{J.}~\bibnamefont{Rothman}},
  \bibinfo{author}{\bibfnamefont{M.}~\bibnamefont{Kl\"{a}ui}},
  \bibnamefont{and} \bibinfo{author}{\bibfnamefont{J.~A.~C.}
  \bibnamefont{Bland}}, \bibinfo{journal}{J.~Appl. Phys.}
  \textbf{\bibinfo{volume}{89}}, \bibinfo{pages}{7579} (\bibinfo{year}{2001}),
  \urlprefix\url{http://link.aip.org/link/?JAP/89/7579/1}.

\bibitem[{\citenamefont{Kl{\"a}ui et~al.}(2003)\citenamefont{Kl{\"a}ui, Vaz,
  Lopez-Diaz, and Bland}}]{Klaui03a}
\bibinfo{author}{\bibfnamefont{M.}~\bibnamefont{Kl{\"a}ui}},
  \bibinfo{author}{\bibfnamefont{C.~A.~F.} \bibnamefont{Vaz}},
  \bibinfo{author}{\bibfnamefont{L.}~\bibnamefont{Lopez-Diaz}},
  \bibnamefont{and} \bibinfo{author}{\bibfnamefont{J.~A.~C.}
  \bibnamefont{Bland}}, \bibinfo{journal}{Journal of Physics: Condensed Matter}
  \textbf{\bibinfo{volume}{15}}, \bibinfo{pages}{R985} (\bibinfo{year}{2003}),
  \urlprefix\url{http://stacks.iop.org/0953-8984/15/R985}.

\bibitem[{\citenamefont{Zhang and Haas}(2010)}]{Zhang10}
\bibinfo{author}{\bibfnamefont{W.}~\bibnamefont{Zhang}} \bibnamefont{and}
  \bibinfo{author}{\bibfnamefont{S.}~\bibnamefont{Haas}},
  \bibinfo{journal}{Phys. Rev. B} \textbf{\bibinfo{volume}{81}},
  \bibinfo{pages}{064433} (\bibinfo{year}{2010}),
  \urlprefix\url{http://link.aps.org/doi/10.1103/PhysRevB.81.064433}.

\bibitem[{\citenamefont{Landeros et~al.}(2006)\citenamefont{Landeros, Escrig,
  Altbir, Bahiana, and d'Albuquerque~e Castro}}]{Landeros06}
\bibinfo{author}{\bibfnamefont{P.}~\bibnamefont{Landeros}},
  \bibinfo{author}{\bibfnamefont{J.}~\bibnamefont{Escrig}},
  \bibinfo{author}{\bibfnamefont{D.}~\bibnamefont{Altbir}},
  \bibinfo{author}{\bibfnamefont{M.}~\bibnamefont{Bahiana}}, \bibnamefont{and}
  \bibinfo{author}{\bibfnamefont{J.}~\bibnamefont{d'Albuquerque~e Castro}},
  \bibinfo{journal}{Journal of Applied Physics} \textbf{\bibinfo{volume}{100}},
  \bibinfo{pages}{044311} (\bibinfo{year}{2006}),
  \urlprefix\url{http://dx.doi.org/10.1063/1.2218997}.

\bibitem[{\citenamefont{Kong et~al.}(2008)\citenamefont{Kong, Wang, and
  Chen}}]{Kong08a}
\bibinfo{author}{\bibfnamefont{D.}~\bibnamefont{Kong}},
  \bibinfo{author}{\bibfnamefont{S.}~\bibnamefont{Wang}}, \bibnamefont{and}
  \bibinfo{author}{\bibfnamefont{C.}~\bibnamefont{Chen}},
  \bibinfo{journal}{Journal of Applied Physics} \textbf{\bibinfo{volume}{104}},
  \bibinfo{eid}{013923} (pages~\bibinfo{numpages}{6}) (\bibinfo{year}{2008}),
  \urlprefix\url{http://link.aip.org/link/?JAP/104/013923/1}.

\bibitem[{mag()}]{magpar}
\emph{\bibinfo{title}{{MAGPAR} finite element micromagnetics package}},
  \bibinfo{note}{developed by Werner Scholz. We used the 0.9 release},
  \urlprefix\url{http://www.magpar.net}.

\bibitem[{\citenamefont{Scholz et~al.}(2003)\citenamefont{Scholz, Fidler,
  Schrefl, Suess, Dittrich, Forster, and Tsiantos}}]{Scholz03a}
\bibinfo{author}{\bibfnamefont{W.}~\bibnamefont{Scholz}},
  \bibinfo{author}{\bibfnamefont{J.}~\bibnamefont{Fidler}},
  \bibinfo{author}{\bibfnamefont{T.}~\bibnamefont{Schrefl}},
  \bibinfo{author}{\bibfnamefont{D.}~\bibnamefont{Suess}},
  \bibinfo{author}{\bibfnamefont{R.}~\bibnamefont{Dittrich}},
  \bibinfo{author}{\bibfnamefont{H.}~\bibnamefont{Forster}}, \bibnamefont{and}
  \bibinfo{author}{\bibfnamefont{V.}~\bibnamefont{Tsiantos}},
  \bibinfo{journal}{Computational Materials Science}
  \textbf{\bibinfo{volume}{28}}, \bibinfo{pages}{366 } (\bibinfo{year}{2003}),
  ISSN \bibinfo{issn}{0927-0256}, \bibinfo{note}{proceedings of the Symposium
  on Software Development for Process and Materials Design},
  \urlprefix\url{http://www.sciencedirect.com/science/article/pii/S0927025603001198}.

\bibitem[{\citenamefont{Metlov and Guslienko}(2002)}]{Metlov02}
\bibinfo{author}{\bibfnamefont{K.~L.} \bibnamefont{Metlov}} \bibnamefont{and}
  \bibinfo{author}{\bibfnamefont{K.~Y.} \bibnamefont{Guslienko}},
  \bibinfo{journal}{J.~Magn. Magn. Mater.} \textbf{\bibinfo{volume}{242-245}},
  \bibinfo{pages}{1015} (\bibinfo{year}{2002}),
  \urlprefix\url{http://www.sciencedirect.com/science/article/B6TJJ-44PKP2R-F/2/3cb3318ef44a490869eeae7cf4d79650}.

\bibitem[{uni()}]{unicc}
\emph{\bibinfo{title}{Kyiv {N}ational {T}aras {S}hevchenko {U}niversity
  high--performance computing cluster}},
  \urlprefix\url{http://cluster.univ.kiev.ua/eng/}.

\end{thebibliography}

%

%

\end{multicols}

\end{document}